\documentclass[manuscript,screen]{acmart}
\usepackage{adjustbox}

\AtBeginDocument{%
  \providecommand\BibTeX{{%
    \normalfont B\kern-0.5em{\scshape i\kern-0.25em b}\kern-0.8em\TeX}}}

\setcopyright{acmlicensed}
\copyrightyear{2024}
\acmYear{2024}
\acmDOI{XXXXXXX.XXXXXXX}

\usepackage{soul}
\usepackage{algorithm, algpseudocode}
\usepackage{graphicx}
\usepackage{subcaption}
\usepackage{multirow}
\begin{document}

\title[Federated Anomaly Detection for Early-Stage Diagnosis of ASD using Serious Game Data]{Federated Anomaly Detection for Early-Stage Diagnosis of Autism Spectrum Disorders using Serious Game Data}

\author{Nikolaos~Pavlidis}
\email{npavlidi@ee.duth.gr}
\orcid{https://orcid.org/0000-0001-9370-5023}
\author{Vasileios~Perifanis}
\email{vperifan@ee.duth.gr}
\author{Eleni~Briola}
\email{ebriola@ee.duth.gr}
\author{Christos-Chrysanthos~Nikolaidis}
\email{cnikolai@ee.duth.gr}
\author{Eleftheria~Katsiri}
\email{ekatsiri@ee.duth.gr}
\author{Pavlos~S.~Efraimidis}
\email{pefraimi@ee.duth.gr}
\affiliation{%
  \institution{Democritus University of Thrace}
  \city{Xanthi}
  \country{Greece}
  \postcode{67100}
}
\author{Despina Elisabeth Filippidou}
\email{elizabeth@dotsoft.gr}
\affiliation{%
  \institution{DOTSOFT SA}
  \city{Thessaloniki}
  \country{Greece}
  \postcode{55535}
}

\renewcommand{\shortauthors}{Pavlidis and Perifanis, et al.}

\begin{abstract}
Early identification of Autism Spectrum Disorder (ASD) is considered critical for effective intervention to mitigate emotional, financial and societal burdens. Although ASD belongs to a group of neurodevelopmental disabilities that are not curable, researchers agree that targeted interventions during childhood can drastically improve the overall well-being of individuals. However, conventional ASD detection methods such as screening tests, are often costly and time-consuming. This study presents a novel semi-supervised approach for ASD detection using AutoEncoder-based Machine Learning (ML) methods due to the challenge of obtaining ground truth labels for the associated task. Our approach utilizes data collected manually through a serious game specifically designed for this purpose. Since the sensitive data collected by the gamified application are susceptible to privacy leakage, we developed a Federated Learning (FL) framework that can enhance user privacy without compromising the overall performance of the ML models. The framework is further enhanced with Fully Homomorphic Encryption (FHE) during model aggregation to minimize the possibility of inference attacks and client selection mechanisms as well as state-of-the-art aggregators to improve the model's predictive accuracy. Our results demonstrate that semi-supervised FL can effectively predict an ASD risk indicator for each case while simultaneously addressing privacy concerns.
\end{abstract}




\maketitle

\section{Introduction}
Current research advancements show that delayed diagnosis of communication disorders and a lack of early intervention can lead to increased emotional and financial costs for both individuals and their families~\cite{franz2022review, perochon2023phenotyping, rabbi2021cnnasd}. Additionally, such issues can place a burden on the state's economic indicators due to heightened needs for special education programs and healthcare services. Although estimating the exact cost of communication disorders on the economy is challenging, studies indicate that it ranges from \$154.3 billion to \$186 billion annually in the US alone~\cite{ruben2000}. To mitigate these costs and improve children's well-being, it is crucial to promptly diagnose communication issues and provide targeted intervention programs in partnership with clinician experts~\cite{ward1999}. 

Communication specialists often rely on psychometric tests to assess a child's cognitive processes, reasoning skills and responses to diagnose communication problems. After collecting data from psychometric tests, Deep Learning (DL) techniques can be employed to analyze the data and compute risk indicators for a child to develop a communication disorder. Given the diverse spectrum of symptoms characteristic of these disorders, incorporating representative data from various sources is vital for robust and accurate predictions.~\cite{cummings2023disorders}. However, centrally collecting and processing health related data raise significant concerns related to data privacy and confidentiality~\cite{sharma2023healthprivacy, mahmud2022xandprivate}. On top of that, different entities and clinicians are often reluctant to share their own private information with third parties.

Federated Learning (FL)~\cite{mcmahan2017fed} emerges as a privacy preserving technique that enables collaborative training of a DL model without revealing private information, thus mitigating the identified issues with data privacy. FL enables distributed machine learning (ML) and edge intelligence in various domains including recommendation systems~\cite{perifanis2022fedncf, perifanis2023fedpoi}, communications~\cite{perifanis20235g}, industry~\cite{hegiste2022fedindustry}, social-care\cite{nikolaidis2023ifmbe} and healthcare~\cite{joshi2022fedhealth, jourdan2020privacyhealth}. It has been proved especially beneficial for health related applications, where user privacy is considered of high importance, by enabling collaboration among different health institutions or individual clinicians.

In this work, we focus on Autism Spectrum Disorders (ASD), a group of neurodevelopmental disorders, often identified in childhood, characterized by deficits in social communication and interaction and the presence of restricted and repetitive behaviors~\cite{american2013diagnostic}. According to recent estimates from the Center for Disease Control and Prevention (CDC) - Autism and Developmental Disabilities Monitoring (ADDM) Network, approximately 1 in 36 children has been identified with ASD~\cite{maenner2023prevalence}.

In response to the need for early ASD identification and targeted interventions that can ease symptoms of the disorder and improve the quality of life of ASD suffering individuals~\cite{camarata2014early}, we developed a software application suite, \textbf{F}ederated \textbf{L}earning \textbf{A}pplication for diagnosis of co\textbf{M}munication disord\textbf{E}rs i\textbf{N} \textbf{C}hild devel\textbf{O}pment (FLAMENCO). Our framework leverages ML in a gamified environment to predict a personalized risk indicator for normal communication skill development in children. Clinicians can use this risk indicator for timely and accurate ASD diagnosis.

For data collection, we utilised a mobile application aimed at monitoring normal development of communication skills in children (aged 3-7) with targeted exercises designed by speech therapists. The application provides a fun gaming experience for children, while guided by their clinician, through various game challenges. Specifically, the system is based on a tablet animation game, where children complete tasks and respond to challenges, with their activities being recorded. These data are then processed and analyzed, generating scores that clinicians can access (with parental consent) to evaluate a child's communication abilities.

\begin{figure}[ht!]
    \centering    
    \includegraphics[width=.9\columnwidth]{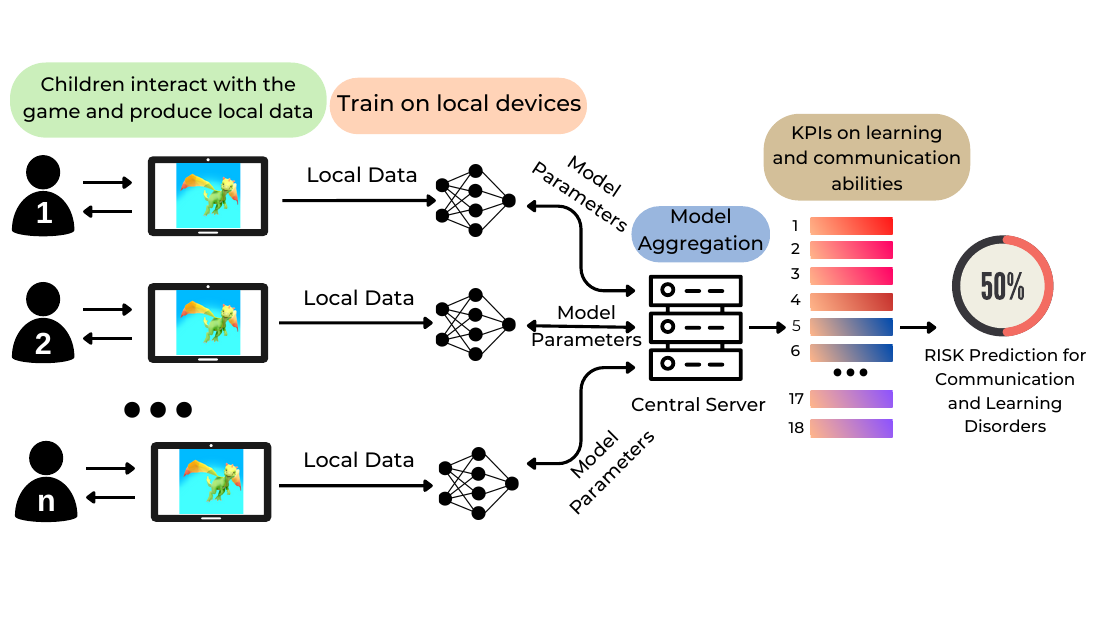}
    \caption{FLAMENCO: A Federated Learning Framework for diagnosis of communication disorders in child development}
    \label{fig:flamenco_overview}
\end{figure}

Since acquiring accurate ground truth labels for our downstream task is rather difficult due to complexity of the ASD development stages~\cite{toki2023}, we opted for a semi-supervised ML solution, modelling the task as anomaly detection. This approach identifies outlier groups from the norm in children's communication skill development. Our framework, designed for federated settings, enhances data privacy and fosters collaboration among clinicians. A high level overview of the FLAMENCO framework is illustrated in Fig.~\ref{fig:flamenco_overview}.

As part of this work, we also identified and experimented with two open issues in FL, namely client selection and securing exchanged weights. To mitigate these issues, we incorporate client selection mechanisms based on client's contribution to the overall learning performance and Fully Homomorphic Encryption (FHE) for enhanced privacy.


The main contributions of this work are summarized as follows:
\begin{itemize}
    \item We introduce a semi-supervised FL framework to identify potential communications disorders related to the autism spectrum in children development. 
    \item We release a novel dataset designed for predicting potential deficiencies regarding ASD, tailored for FL applications~\cite{flamenco_dataset}.
    \item We experimentally evaluate eight popular aggregators to address the problem of statistical heterogeneity.
    \item We utilize FHE during model aggregation to enhance data privacy.
    \item We propose a novel client selection mechanism that improves the robustness and learning stability under FL settings.
\end{itemize}

The rest of this paper is structured as follows: Section~\ref{related_work} introduces the main concepts used in this paper regarding applications of FL for identifying communication disorders in child development and reviews the related work. In Section~\ref{datasets}, we provide an in depth presentation of the dataset we publicly release, as well as a similar dataset from the literature for comparison reasons. Section~\ref{problem_formulation} defines the settings of our experiments, the utilized algorithms and evaluation metrics, as well as the techniques we employ to enhance privacy, robustness and generalization ability of our framework. Section~\ref{experiments} discusses the experimental results and showcases the applicability of our framework in real-world environments. Finally, Section~\ref{conclusion} concludes our work and proposes further research directions.

\section{Preliminaries and Related Work}
\label{related_work}
In this section, we present the ML concepts utilized in this work and how our proposed solution differentiates from the related work in the domain.

\subsection{ASD Datasets}
Early diagnosis of ASD can be very challenging, especially when considering the young age of the examined patients. For this reason, several approaches are being followed by the clinicians to identify ASD traits~\cite{falkmer2013diagnostic}. The rise of ML applications in healthcare, has been proved an important tool that fosters early detection of ASD cases by analyzing their behavioral and clinical data~\cite{hyde2019applications}. Examples of ML applications in autism patterns recognition include speech analysis~\cite{pahwa2016machine}, eye tracking~\cite{jeyarani2023eye}, facial expression and behavior recognition through video analysis~\cite{wu2021machine, liu2016identifying} and image analysis of MRI or EEG scans~\cite{sharif2022novel}. The most prevalent applications of ML in the related literature though leverage tabular data from screening tests. For example, Abbas et al.~\cite{abbas2018machine} utilized the popular Autism Diagnostic Interview Revised (ADI-R) and Autism Diagnostic Observation Schedule Revised (ADOS-R) screening tests and introduced a new evaluation tool with implementation of different attribute encoding approaches to resolve data insufficiency, non-linearity, and inconsistency issues. Allison et al.~\cite{allison2012toward} proposed a short quantitative checklist that can be used at several stages of the life of a patient, including toddlers, children, adolescents and young adults. Similarly, Thabtah et al.~\cite{thabtah2018new} developed a mobile application named ASDTests based on Q-CHAT and AQ-10 tools that help in the early identification of ASD patterns. Later, the data gathered using ASDTests application were made publicly available through Kaggle and the University of California-Irvine (UCI) Machine Learning (ML) repository as open source datasets, thus facilitating researchers' interest on the topic. There are four available ASDTests datasets for different age ranges (Toddlers, Children, Adolescents and Adults) comprising of patients' demographic data, as well as their answers to ten (A1-A10) binary questions. Based on each case's answers, an indicative score is calculated by the app which ranges from 0 to 10, with a final individual score of more than 6 out of 10 indicates a positive prediction of ASD.

\subsection{Machine Learning for ASD Detection}
ML has been widely used on ASD screening data for early and precise identification of abnormal traits. Omar et al.~\cite{omar2019machine} developed an autism prediction model by merging two decision tree based models, i.e. Random Forest-CART and Random Forest-Id3, improving the overall predictive accuracy of the model. Hossain et al.~\cite{hossain2021detecting} evaluated state-of-the-art classification and feature selection techniques to determine the best performing classifier and feature set, respectively, for the four ASDTests datasets. They suggest that a combination of Multi-Layer Perceptron and relief-F feature selection techniques is optimal for all four datasets achieving 100\% accuracy with minimal number of attributes. Hasan et al.~\cite{hasan2022machine} employed several ML techniques alongside four different Feature Scaling (FS) strategies i.e., Quantile Transformer (QT), Power Transformer (PT), Normalizer, and Max Abs Scaler (MAS) using the ASDTests datasets. Their results show that AdaBoost (AB) and Linear Discriminant Analysis (LDA) classifiers rank higher among others when combined with Normalizer and QT scaling techniques, respectively. Thabtah and Peebles~\cite{thabtah2020new} proposed a new ML technique called Rules-Machine Learning (RML) that offers users a knowledge base of rules for understanding the underlying reasons behind the classification in addition to detecting ASD traits. Finally, Akter et al.~\cite{akter2019machine} evaluated feature transformation methods including log, Z-score and sine functions for the four datasets, demonstrating that gradient boosting algorithms have a better overall performance when combined with the Z-score transformation.

Considering related work on datasets coming from serious games, Toki et al.~\cite{toki2023applying} employed Artificial Neural Networks (ANNs), K-Nearest Neighbors (KNN), Support Vector Machines (SVM), along with state-of-the-art Optimizers, namely the Adam, the Broyden–Fletcher–Goldfarb–Shanno (BFGS), Genetic algorithms (GAs) and the Particle Swarm Optimization algorithm (PSO) in a dataset similar to the one presented in this paper, concluding that Integer-bounded Neural Network proved to be the best competitor.

It worth mentioning that the previously discussed works treat ASD identification as a supervised ML classification problem. However, this approach has significant limitations in real-world scenarios, as it requires the availability of ground truth labels for training the model. In practical applications, especially concerning ASD with its varying developmental stages, obtaining these labels is challenging due to insufficient number of available diagnoses to effectively train a ML model. To the best of our knowledge, our proposed solution is the first work that addresses the problem of ASD identification as a semi-supervised anomaly detection problem under FL settings. This approach allows ML models to learn from a subset of available data consisting of typical development cases. Specifically, it optimizes the model to learn their distribution, effectively leading to detecting potential anomalies that could indicate a high risk of ASD development.

\subsection{Anomaly Detection}
Anomaly detection is a widely used technique that enables the identification and extraction of anomalous data points from a given data distribution. Early algorithms that were utilized for anomaly detection was statistical based models~\cite{chandola2010anomaly}. These methods build a statistical model on the provided data distribution and during inference, they can detect whether or not an instance belongs to this model. Several methods are used to conduct statistical anomaly detection including proximity based, parametric, non-parametric and semi-parametric methods~\cite{bhuyan2013network}. Recent developments in ML though, have found direct applications on anomaly detection problem. The most common approach handles anomaly detection as a semi-supervised ML problem. Semi-supervised techniques presume that the training distribution represents only the normal class. Thus, any data point that does not match the given distribution is considered as anomalous. Since they do not need the associated class labels for each input feature, semi-supervised methods are more common than the supervised ones~\cite{nassif2021machine}. Anomaly detection finds applications in various domains such as intrusion detection systems \cite{hajj2021anomaly}, networks~\cite{ahmed2016survey} and time-series~\cite{blazquez2021review}. 

In our proposed approach, we utilize an AutoEncoder model to execute anomaly detection, focusing on the model's reconstruction loss. This method requires  class labels for only a subset of normal data points. These labeled points then serve as the training set for the model, thus reducing the need for the time-consuming manual labeling.

\subsection{Federated Learning}
Federated Learning is a distributed machine learning technique, originally proposed by McMahan et al.~\cite{mcmahan2017fed}, which enables collaborative training of a ML model across multiple clients, while preserving data privacy. During the FL process, each participating client performs local training on its private data using the model weights shared by a central server that orchestrates the whole process. Then, in contrast to centralized machine learning, participating entities only share their calculated model parameters. The central server receives the parameter updates of each client and performs an aggregation operation. The aggregation result corresponds to the new global model, which is then transmitted to the participants for the next FL round. FL's market value is expected to double by 2030~\cite{PolarisReport}, however there are still various open problems need to be addressed~\cite{kairouz2021advances}. For instance, some works~\cite{zhang2020batchcrypt, hijazi2023secure} question the security of the exchanged model parameters, highlighting that potential attacks during this process could lead to private information leakage. To mitigate this problem, they propose to use tools such as FHE, Secure Multi-Party Computation and Differential Privacy as privacy-enhancements to FL.

Another major challenge for FL, as mentioned in~\cite{zhu2021noniid}, is handling non independent and identically distributed (non-IID) data. Potential solutions to this issue focus on designing client selection mechanisms that favors clients with higher contribution to the overall performance~\cite{nishio2019client} or using functions during the model weights aggregation step that specialize in handling non-IID distributions~\cite{qi2023aggregation}.

In this work, we utilize FHE as a privacy-enhancement during the parameters aggregation step. Moreover, we evaluate our framework with eight (8) different state-of-the-art aggregators and four (4) client selection mechanisms, aiming to assess the robustness of our method and improve its overall predictive accuracy, while maintaining low communication cost.

\subsection{Federated Learning for ASD Identification}
Various attempts have been made for privacy-oriented frameworks to identify potential learning disabilities in children development, since the data associated with these applications are considered highly confidential. Shamseddine et al.~\cite{shamseddine2022feasibility} proposed a privacy-preserving model to predict ASD based on individual's behavioral traits or facial features. Farooq et al.~\cite{farooq2023detection} employed FL using ASDTests datasets, having a predictive accuracy of 98\% in children cases and 81\% in adults. Finally, Lakhan et al.~\cite{lakhan2023autism} implemented a FL framework that allows collaborative training of a ML model with heterogeneous sources of data from different ASD laboratories.

Our proposed framework is the first that evaluates the use of FL on serious game data for ASD diagnosis, offering a practical scenario where children can complete the screening tests easily through a gamified application on their smart device and then get fast and accurate risk prediction results without compromising their privacy.

\section{Dataset}
\label{datasets}
In the context of this work, we utilized data derived from a serious game application similar to the one presented in~\cite{toki2021game}. Toki et al. developed a mobile application that assesses children communication development through specialised gamified exercises that have been designed with the assistance of expert clinicians. The application monitors children's performance on various tasks and computes communication skills indicators, with scores ranging between 0-100. These indicators can be used as input for a ML model to predict the risk for each case to develop communication disorder symptoms.

Similar to ~\cite{toki2021game}, we collected, processed and publicly released an anonymized dataset~\cite{flamenco_dataset} that derived from childrens' interactions with the game application. The dataset was specially designed for federated settings by co-operating with clinicians that introduced the application to their cases. As we mainly targeted ASD, we utilised some of the available communication skills indicators as presented in Table~\ref{indicators}.

\begin{table}[ht!]
\caption{Communication Indicators List for the FLAMENCO dataset.}
\label{indicators}
\begin{tabular}{|c|p{10cm}|}
\hline
\textbf{Indicator} & \textbf{Description} \\ \hline
Verbalization & The ability to express oneself using words.\\ \hline
Voicing & Speaking at an appropriate volume and tone.\\ \hline
Articulation & Pronunciation and clarity in speech.\\ \hline
Phonology & Understanding and using sounds in language.\\ \hline
Syntax & The arrangement of words and phrases in a sentence\\ \hline
Perception & Awareness and understanding of sensory inputs\\ \hline
Fine Motor & Small muscle movements for activities like writing.\\ \hline
Pre-writing & Skills necessary before formal writing begins. Includes activities like drawing shapes, patterns, and lines.\\ \hline
Visual-motor integration & Coordination between visual perception and motor skills.\\ \hline
Spatial Orientation & Understanding of space, direction, and position.\\ \hline
Sequencing & Organizing information or actions in a logical order.\\ \hline
Memory & Retaining and recalling information. \\ \hline
Recognition & Identifying and naming objects or patterns. \\ \hline
Cognitive Flexibility & Ability to adapt and switch between tasks or thoughts. \\ \hline
\end{tabular}
\end{table}

Except for the respective scores per communication ability each row also contains the following details:
\begin{itemize}
    \item case\_id: An anonymized identifier used to distinguish cases.
    \item client\_id: Identifies the client to which the case belongs, useful for dataset splitting in federated settings.
    \item target: Can be -1 (no clinician's diagnosis available for the case), 0 (no diagnosed deficiency in the case), 1 (indicates a positive diagnosis of communication deficiency by a clinician)
\end{itemize}

Aiming to prove our framework's generalization ability and robustness of our results, we further utilized ASDTest Dataset (Children)~\cite{thabtah2018new}, a commonly used dataset in the literature. Since we used this dataset only for verification reasons, we utilized all the available variables without excessive data preprocessing. For pre-processing, the categorical features were transformed with one-hot label encoder and normalized using the Min-Max normalization. The target column in this dataset contains only zeros (no diagnosed deficiency) and ones (diagnosed autism disorder), without any unidentified cases (-1). In order to make the dataset compatible with our FL framework we equally distributed the cases into five potential clients, i.e., clinicians.

\begin{figure}[ht!]
    \centering    
    \includegraphics[width=\columnwidth]{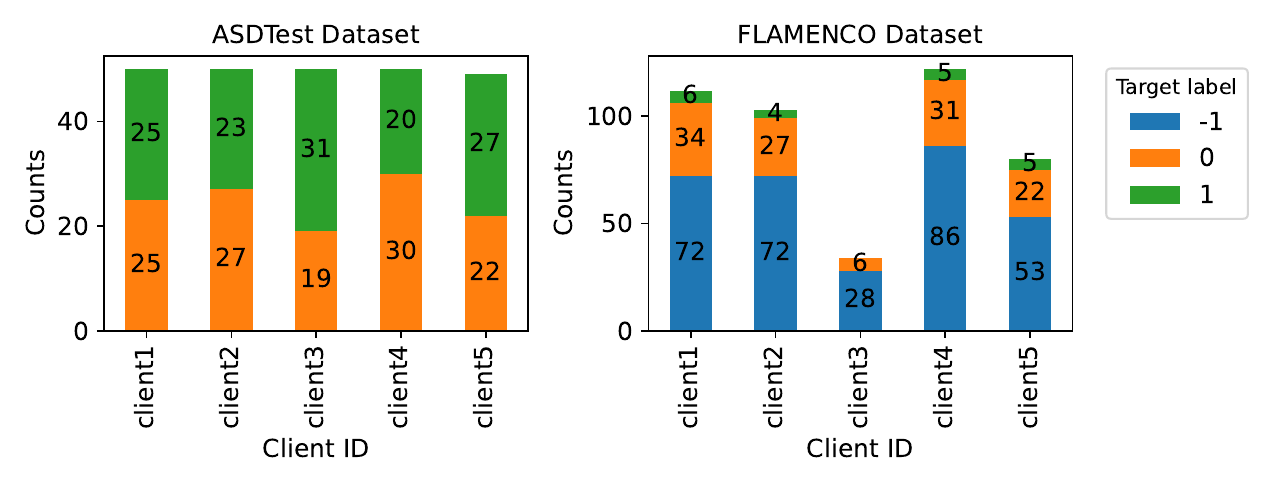}
    \caption{Label distribution among clients.}
    \label{fig:targets}
\end{figure}

It is important to highlight that in the FLAMENCO dataset, the cases are not equally distributed among the available clients, simulating a more realistic scenario. As presented in Fig.~\ref{fig:targets}, the label distribution in FLAMENCO presents both quantity and label skew, meaning that some clients may not have a representative case for each label. On the other hand, the ASDTest dataset is more balanced with all clients having both normal and abnormal cases. Table \ref{datasets_tab} presents a high level view of dataset statistics used in this work.

\begin{table}[ht!]
\caption{Datasets Overview}
\label{datasets_tab}
\begin{tabular}{c|c|c|c|c|c}
Dataset           & Total Cases & Train Cases & Test Cases & Number of Features & Number of Clients \\ \hline
FLAMENCO Dataset  & 451         & 192         & 259        & 19                 & 5               \\
ASDTests Dataset & 249         & 98         & 151         & 20                 & 5               \\ \hline
\end{tabular}
\end{table}
\begin{figure}[ht!]
    \centering    
    \includegraphics[width=\columnwidth]{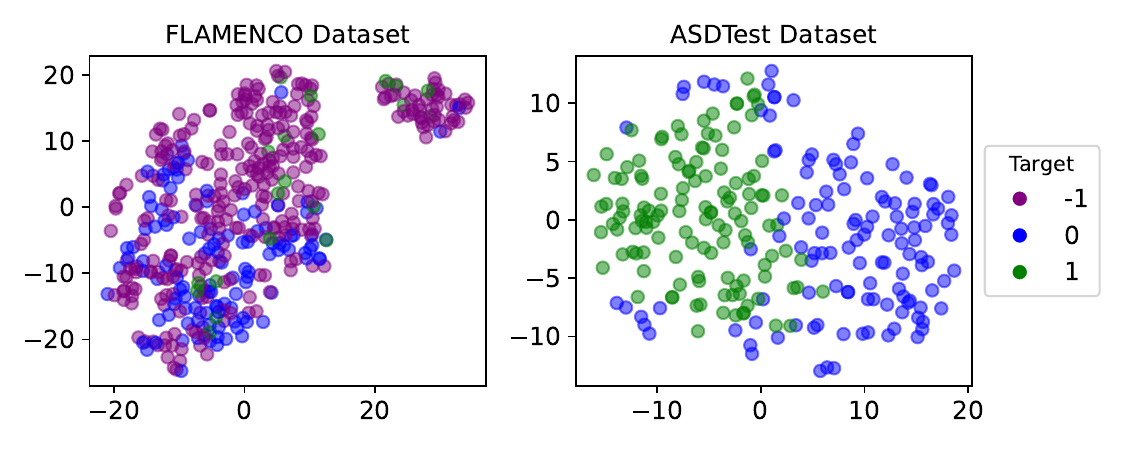}
    \caption{TSNE Representation of the target column for considered datasets.}
    \label{fig:tsne_datasets}
\end{figure}
Fig.~\ref{fig:tsne_datasets} illustrates a TSNE visualization for the two  datasets. The visualization for the FLAMENCO dataset suggests that the cases are not easily separable in the reduced feature space. More precisely, a significant overlap between the different labels is observed. In addition, the clusters do not show a clear separation, which implies that there is not a straightforward linear boundary that separates between negative and positive deficiency diagnosis. In contrast, the cases in ASDTest dataset seem present higher degree of separability, indicating that ML models could easier fit the training samples.

\section{Problem Formulation and Methods}
\label{problem_formulation}
In this section, we present our approach to handle the problem of ASD detection as a semi-supervised machine learning task modeling it as anomaly detection under FL settings.

\subsection{Federated Anomaly Detection}
The main downstream task associated with the datasets presented in the previous sections, involves predicting the probability of a case being diagnosed with ASD. To this end, we utilize AutoEncoder-based architectures, a type of neural networks that try to reconstruct the original input. The goal throughout experiments was to train the model on “normal” data and learn a hidden representation of the normal distribution. When the network encounters data that significantly differ from the training samples, i.e. cases with ASD, we anticipate higher reconstruction error since the associated patterns differ from what the model learned as normal. During inference, when the model encounters samples that are hard to reconstruct accurately, it will result in high reconstruction error, indicating a potential disability.

The datasets train-test split was performed based on our proposed solution architecture that handles this problem as an anomaly detection. In other words, the training set in each case comprises data points that are considered as normal and are labeled as 0 (no diagnosed deficiency) and specifically for the FLAMENCO dataset with some selected unlabeled cases (-1) that are presumed to be normal. The inclusion of unlabeled cases in the training set is based on the assumption that normal instances are much more common than anomalies in a real-world scenario \cite{nassif2021machine} and are found based on high scores achieved through the gamified environment. During training, the model is evaluated on all possible labels, i.e., negative, unknown and positive cases.

Since privacy is considered of high importance for our case study scenario, involving sensitive medical data for each children, FL is favored against a centralized solution. On both datasets we have the total cases distributed in 5 clients, equivalent to the clinicians that are responsible for the cases. A central server is orchestrates the communication among clients and the FL process in general. It starts every federated round by transmitting the learnable weights of a common global model to clients. Then, each client performs local training on its private data using gradient descent optimization. After local training, every client re-transmits its local model weights back to the central server, which aggregates them to produce the new global weight. The whole process is described in Algorithm \ref{alg:alg1}.

\begin{algorithm}[ht!]
\caption{Federated Anomaly Detection for Early ASD identification with Serious Game Data.}\label{alg:alg1}
\textbf{Input:} Clinicians $Cs = \{C_1, C_2, ..., C_n\}$. $\mathcal{R}$ is the number federated rounds, $E$ is the number of local epochs, $B$ is the batch size, $\eta$ is the learning rate and $\nabla \mathcal{L}$ is the gradient optimization objective.\\
\textbf{Output:} Model weights $w$. 
\begin{algorithmic}[1]
\State Initialize $w_0$.
\For{each round $r = 1, 2, ..., \mathcal{R}$}
    \State $\{C_r\} \leftarrow$ sample round participants from $Cs$.
  \State Transmit global $w_{r-1}$ to each clinician $k \in \{C_r\}$
  \For{each clinician $k \in \{C_r\}$ \textbf{in parallel}}
    \State $w^k_{r} \leftarrow$ LocalTraining($k$, $w_{r-1}$)
  \EndFor
    \State $w_{r+1} \leftarrow Agg(\bigcup_{k \in \{K_r\}}w_{r}^k$)
\EndFor  
\Function{LocalTraining}{$k, w$} 
    \State dataset $\leftarrow$ Select subjects without deficiency.
    \State $\mathcal{B} \leftarrow$ split local dataset into batches of size $B$.
    \For{each local epoch $e=1, ..., E$}
        \For{batch $b \in \mathcal{B}$}
            \State \hspace{0.5cm}$w \leftarrow w - \eta \nabla \mathcal{L}(w;b)$
        \EndFor
    \EndFor
\State \textbf{return} $w$ to server.
\EndFunction
\end{algorithmic}
\label{alg1}
\end{algorithm}

\subsection{Models}
As previously stated, our proposed anomaly detection solution is based on measuring the reconstruction error of the AutoEncoder-based ML models that have been fitted on data considered as normal. We utilize two common model architectures, namely, an AutoEncoder and a Variational AutoEncoder.

\textbf{AutoEncoder}: An AutoEncoder \cite{kramer1991ae} is a type of machine learning model that learns to compress data into a reduced hidden representation and then reconstructs the original data as accurately as possible from this compressed form. It consists of an encoder that transforms the input into a latent space and a decoder that reconstructs the input from this representation. This characteristic makes AutoEncoders particularly suitable for our task, where the model is trained on normal cases. The model learns a hidden representation of these normal cases and we expect a low reconstruction error for negative diagnoses, enabling the detection of positive instances by highlighting them with higher reconstruction error. In this work, the encoder consists of two linear layers each with 64 units and a dropout layer between them for regularization. The decoder has the exact opposite architecture of the encoder and in the last layer it applies a sigmoid function to generate an output between 0 and 1.

\textbf{Variational AutoEncoder (VAE)}: A Variational Autoencoder (VAE) \cite{kingma2019vae} is an extension of an AutoEncoder that not only compresses data into a latent space but also learns the statistical properties of the data distribution. It does this by incorporating a probabilistic approach, enabling it to generate new data points by sampling from the learned distribution in the latent space, making it adept at generating diverse and realistic outputs. The architecture of the selected VAE is similar to the one presented in the AutoEncoder model, but with two additional linear layers after the encoder part. These two layers enable the calculation of mu and logvar, which output the parameters for the Gaussian distribution from which we sample the latent representation. Similar to the AutoEncoder, we expect that positive diagnoses will have a higher reconstruction error than negative ones.

\subsection{Evaluation Metrics}
Evaluating anomaly detection algorithms is a challenging task due to the absence of ground truth labeled data \cite{campos2016evaluation}. Nevertheless, there are some commonly accepted classification-based metrics to estimate the model’s predictive performance. In this work, we use two ranking-based metrics to assess our models ability to separate between negative and positive deficiency diagnosis along with one metric that does not require labelled data.

\textbf{Area Under the Receiver Operating Characteristics Curve (AUC-ROC)}: This metric takes into account the entire outlier ranking and is used to measure how well the anomaly detection system is capable of separating the classes of normal and anomalous instances. A high ROC AUC score indicates that, in the overall ranking, outliers are more likely to be ranked ahead of inliers, but it does not necessarily mean that the top rankings are dominated by outliers.

\textbf{Average Precision (AP)}: Average precision is a way to summarize the precision-recall curve into a single value representing average precision across recall levels. In the context of anomaly detection, higher AP scores represent higher quality predictions.

\textbf{Similarity-based Internal, Relative Evaluation of Outlier Solutions (SIREOS)} \cite{marques2022similarity}: This metric is a recent advancement in the domain of evaluating unsupervised algorithms in the absence of ground truth labels. SIREOS uses a similarity measure to discover how distinct the detected outliers (in the project’s score, positive deficiency diagnosis) are, relative to the rest of the data points. This metric is proposed to overcome the computational demands of IREOS \cite{marques2020internal}, while showing comparable or even superior quality to similar unsupervised metrics. In the context of anomaly detection, lower SIREOS values indicate high quality model predictive performance.

\subsection{Aggregators}
The mathematical function used by the central server for the aggregation of the collected learnable weights, referred to as Aggregator, is a key component in every FL system. In this work, we implemented 8 popular aggregators, i.e. FedAvg~\cite{mcmahan2017fed}, FedNova~\cite{wang2020fednova}, FedAvgM~\cite{hsu2019fedavgm}, FedAdagrad, FedYogi, FedAdam~\cite{reddi2020adaptive}, SimpleAvg, MedianAvg, and evaluated our framework to assess their performance.

FL commonly relies on the FedAvg algorithm, initially introduced by McMahan et al. \cite{mcmahan2017fed}. FedAvg aggregates client models by weighted averaging, giving more weight to clients with larger datasets. However, in certain scenarios, FedAvg might cause objective inconsistency. This happens when the global model converges to a point of a different objective function than the true global one due to non-iid and heterogeneous data. To tackle this issue, alternative methods have emerged to better accommodate data diversity.

Addressing non-iid data challenges, Wang et al. proposed FedNova \cite{wang2020fednova}, which normalizes local model updates during aggregation. FedNova normalizes local gradients before aggregation by dividing them by the number of client steps, rather than directly averaging cumulative local gradients without normalization.
FedAvgM, introduced by Tzu et al. \cite{hsu2019fedavgm}, is a variation of FedAvg that utilizes server momentum. This involves iteratively multiplying prior model updates with a hyper-parameter $\beta$ during each epoch, in addition to incorporating new updates.
Furthermore, efforts have been made to create federated versions of adaptive optimizers—FedAdagrad, FedYogi, and FedAdam—as presented in \cite{reddi2020adaptive}. These methods aim to handle heterogeneous data, boost performance, and minimize communication costs.

\subsection{Client Selection Mechanisms}
Under FL setting, managing communication costs is a crucial aspect of the system’s efficiency. Thus, it is essential to implement selection strategies to minimize communication overhead without compromising the predictive performance of the underlying model. To this end, when McMahan et al. \cite{mcmahan2017fed} introduced FL, they utilized a random selection technique that in every federated round sampled a subset of the available clients to participate in the federated process. Later, more sophisticated client selection mechanisms emerged to mitigate even more open problems, such as non-IID datasets, fairness and robustness of the algorithm \cite{fu2023clientselection, pavlidis2023intelligent}.

Towards this goal, we implemented four client selection mechanisms aiming to evaluate how they affect the overall predictive performance of the ML models, as well as their convergence speed and robustness:

\textbf{Random Sampler}:  This selection algorithm involves the server randomly selecting a subset from the available clients to participate in each federated round. This is the first and the simplest client selection mechanism that is being utilized in most FL applications.

\textbf{Std Sampler}:  This algorithm selects clients based on the variability of their data, measured by the standard deviation. Clients are chosen inversely proportional to their standard deviation, i.e., clients with higher variability are less likely to be selected. This sampling technique aims to increase homogeneity of data distribution among selected clients.

\textbf{Quantity Sampler}:  In this selection mechanism, clients are chosen according to the quantity of their samples relative to the number of samples of the overall dataset. This method selects clients with a larger number of samples, thereby penalizing those with fewer samples. This sampler is based on the assumption that clients with fewer samples are less informative in general, so we can select them fewer times in order to decrease communication overheads.

\textbf{Contribution Score Sampler}: Finally, we introduce a novel client selection mechanism that determines a contribution score of each client to the overall performance of the global model and performs the selection of clients based on the ascending order of the clients' contribution score. The contribution of each client is based on clients’ training loss and L2 (or Euclidean) norm of client model divergence from the global and is calculated as follows:

\begin{equation}
    ClientScore = \alpha \sum_{k \in D_{train}} Loss(k) + \beta \left \| w^{client} - w^{global} \right \|_2,
\end{equation}

where $D_{train}$ is the train dataset of the respective client, $w^{client}$ is the vector of learnable weights of client's model after training on local data and $w^{global}$ is the vector of the learnable weights of global model of the previous round. The parameters $\alpha, \beta$ denote the importance of each of the two factors of the score. In our experiments, we used analogies of {50/50, 60/40, 40/60}. Lower $ClientScore$ indicates a homogeneous client with respect on both data distribution and model weights, indicating that this client's participation to the FL process can improve the performance and efficiency of the learning model. Thus, in every FL round after calculating the $ClientScore$ for each participating client, we select those with the lowest scores.

\subsection{Fully Homomorphic Encryption}
FL inherently avoids the need to share raw data. However, there is a risk that an adversary might deduce sensitive information from the model's weights that are directly related to clients’ original data. In our work, we assume that the server might attempt to extract such information. To secure client data against such threats, we have incorporated a FHE mechanism. This allows the server to combine the weights from different clients without accessing the actual weights, thus enabling a blind aggregation. We opted for FHE instead of partial homomorphic encryption (PHE) because FHE supports a broader range of computations, enabling compatibility with various aggregation algorithms. To this end, we employed the TenSEAL homomorphic encryption library \cite{benaissa2021tenseal} and utilized the CKKS scheme \cite{cheon2017homomorphic} for homomorphic computations.

\section{Experimental Evaluation}
\label{experiments}
In this section, we present the results of the experimental evaluation of our proposed framework. For the optimization of the training process we utilized Adam~\cite{kingma2014adam} optimizer with learning rate of 0.001 and set the batch size at 32 samples/batch. In centralized settings we employed 100 training epochs for the FLAMENCO dataset and 200 for the ASDTest Dataset. In federated settings we performed 100 federated rounds for the FLAMENCO dataset and 200 for the ASDTest dataset and in each FL round every participating client performed 3 epochs of local training in both cases. These number of epochs/fl rounds have been empirically demonstrated high quality prediction.
The experiments were conducted on a workstation running Ubuntu 20.04 with 128 GB memory and a RTX 3060 GPU. To validate the accuracy and stability of our results we utilised ten different seeds to initialize random generators and we report the mean values of the respective experiments.
We publicly share the source code of our framework on Github\footnote{\url{https://github.com/nikopavl4/FLAMENCO-Project}} to facilitate further research and reproducibility.

\subsection{Learning Setting Comparison}
In order to validate the applicability and potential of FL for the downstream task, we compare the performance of the ML models under three different learning settings, i.e., traditional centralized learning, individual (or local) learning and federated learning. For the FL setting, we sample all clients per federated round and use the FedAvg aggregation algorithm. The averaged results regarding AUC-ROC, SIREOS and AP scores for the two datasets are reported in Table \ref{allmetrics}. We focus on the AUC-ROC score since the SIREOS and AP scores present similar trends.

\begin{figure}[b!]
    \centering    
    \includegraphics[width=.8\columnwidth]{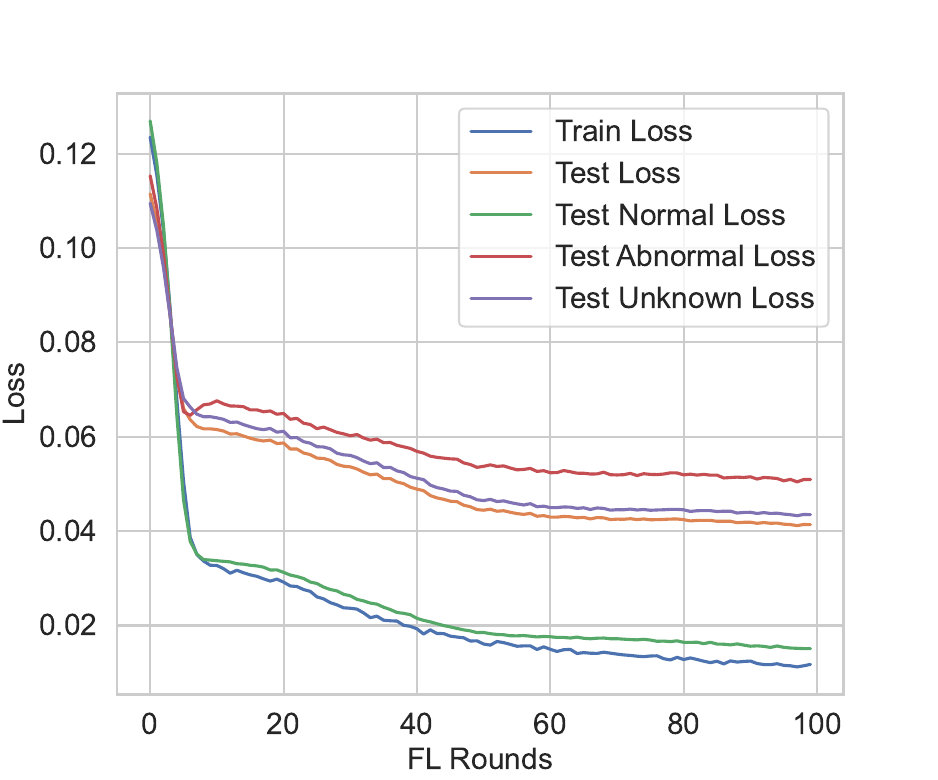}
    \caption{Train and Test Losses (Total \& per Label).}
    \label{fig:losses}
\end{figure}

For both datasets and for the two considered model architectures, the evaluation metrics do not differentiate a lot among the learning settings. Specifically for the FLAMENCO dataset, the \textit{centralized setting} of AutoEncoder presents the highest AUC-ROC score of 79.59\%. \textit{Federated Setting} has a comparable predictive performance of 78.02\% regarding AUC-ROC score. The 2\% decrease compared to the centralized counterpart can be attributed to the differentiation in clients' local distributions and the fewer number of samples per client compared to centralized counterpart. Notably, in \textit{individual setting}, where all clients perform only local training on their private data without any degree of collaboration, the results show accuracy decrease with an AUC-ROC score of 77.43\%. This outcome favors FL as an option that can guarantee high predictive performance of the ML models while maintaining data privacy. Another thing that justifies the efficacy of the proposed federated anomaly detection framework is presented in Fig.~\ref{fig:losses} that shows the loss convergence for each respective label in the test set. As expected, normal cases have lower loss compared to abnormal and unknown ones, showcasing that the model training has effectively captured the patterns of the data points considered as normal.

\begin{table}[t!]
\centering
\caption{Averaged AUC-ROC, SIREOS \& AP scores for all learning settings and ML model types.}
\label{allmetrics}
\begin{adjustbox}{width=\textwidth,center}
\begin{tabular}{cc|ccc|ccc|ccc}
                             &       & \multicolumn{3}{c|}{Centralized} & \multicolumn{3}{c|}{Individual} & \multicolumn{3}{c}{Federated} \\ \hline
\multicolumn{1}{c|}{Dataset} & Model & AUC-ROC    & SIREOS   & AP       & AUC-ROC   & SIREOS   & AP       & AUC-ROC   & SIREOS  & AP      \\ \hline
\multicolumn{1}{c|}{\multirow{2}{*}{FLAMENCO}} & AutoEncoder & 0.7959 & 0.0477 & 0.7451 & 0.7743 & 0.0454 & 0.7399 & 0.7802 & 0.0489 & 0.7422 \\
\multicolumn{1}{c|}{}        & VAE   & 0.7556     & 0.0513   & 0.7298   & 0.7505    & 0.0502   & 0.7103   & 0.7514    & 0.050   & 0.7304  \\ \hline
\multicolumn{1}{c|}{\multirow{2}{*}{ASDTest}}  & AutoEncoder & 0.9806 & 0.1638 & 0.9956 & 0.9417 & 0.1343 & 0.9844 & 0.9698 & 0.1355 & 0.9930 \\
\multicolumn{1}{c|}{}        & VAE   & 0.8726     & 0.1702   & 0.9724   & 0.8522    & 0.1504   & 0.9579   & 0.6403    & 0.1853  & 0.8882  \\ \hline
\end{tabular}
\end{adjustbox}
\end{table}

Similarly, the AutoEncoder demonstrates similar performance on the ASDTest Dataset. Centralized setting performs the best with an AUC-ROC score of 98.06\%, which aligns with the results presented in studies utilizing the ASDTest dataset~\cite{hossain2021detecting, akter2019machine}. This finding supports the potential of using anomaly detection as an effective approach, resulting in predictive performance that is on par with traditional classification tasks, particularly in scenarios with limited ground truth label information. In addition, the federated setting outperforms the individual model with an AUC-ROC score of 96.98\% compared to 94.17\%.

VAE model exhibited inferior performance for all learning settings in both datasets, that in some cases reached even 30\% compared to AutoEncoder.

\subsection{Federated Learning Convergence}
\begin{figure}[t!]
    \centering    
    \includegraphics[width=.9\columnwidth]{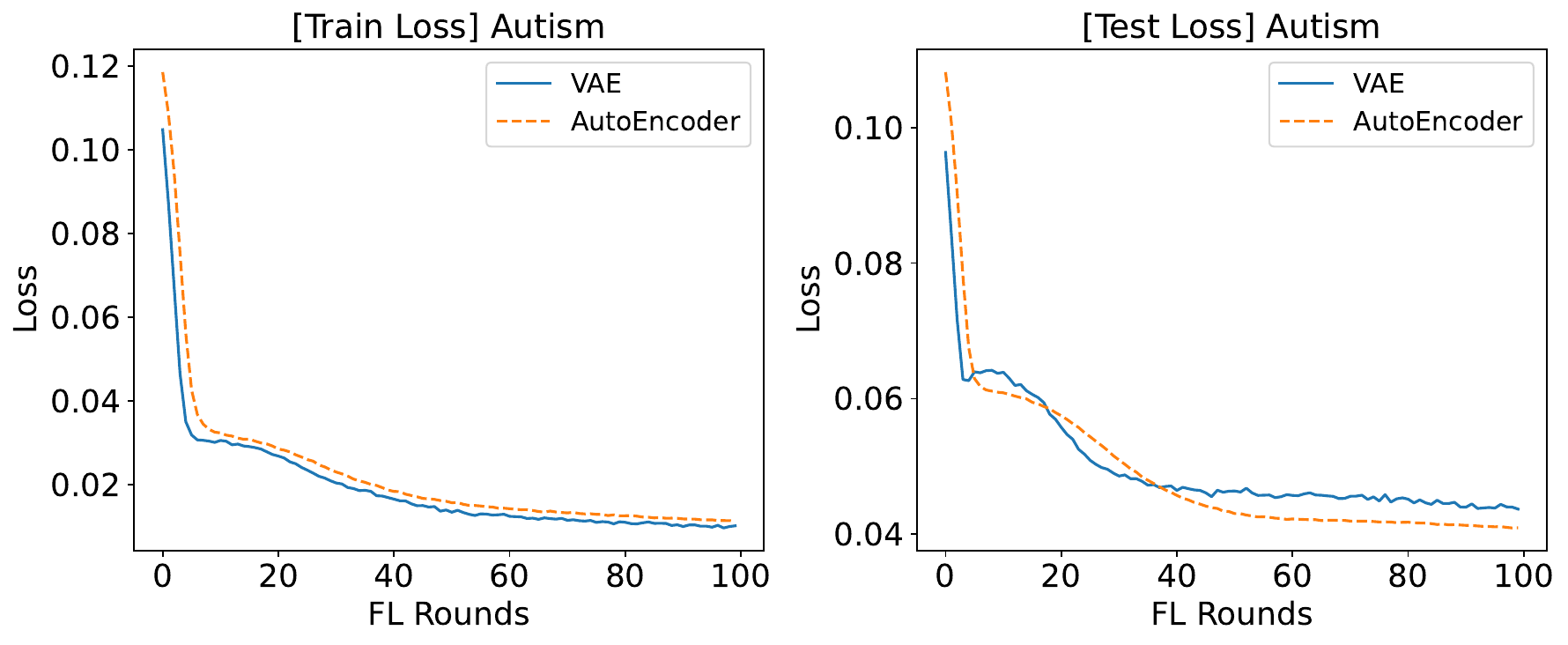}
    \caption{Learning Curves for different ML models for Train and Test}
    \label{fig:aevsvae}
\end{figure}
As a next step, we evaluate the convergence of FL using the two ML model architectures. Fig. \ref{fig:aevsvae} shows the loss curves for the AutoEncoder and VAE models for the training and test sets. As it can be observed, while the VAE model performs slightly better on the training set, it has higher loss (lower accuracy) than the AutoEncoder. In addition, VAE presents fluctuations across learning epochs. On the other hand, the AutoEncoder models not only maintains learning stability but also achieves high performance on both the training and test sets.

Due to the constant underperfomance of VAE and the fact that in general tends to have similar behavior with the AutoEncoder, we will consider only results for the AutoEncoder for the rest of the paper.

\subsection{Aggregators}
\begin{figure}[ht!]
    \centering    
    \includegraphics[width=.9\columnwidth]{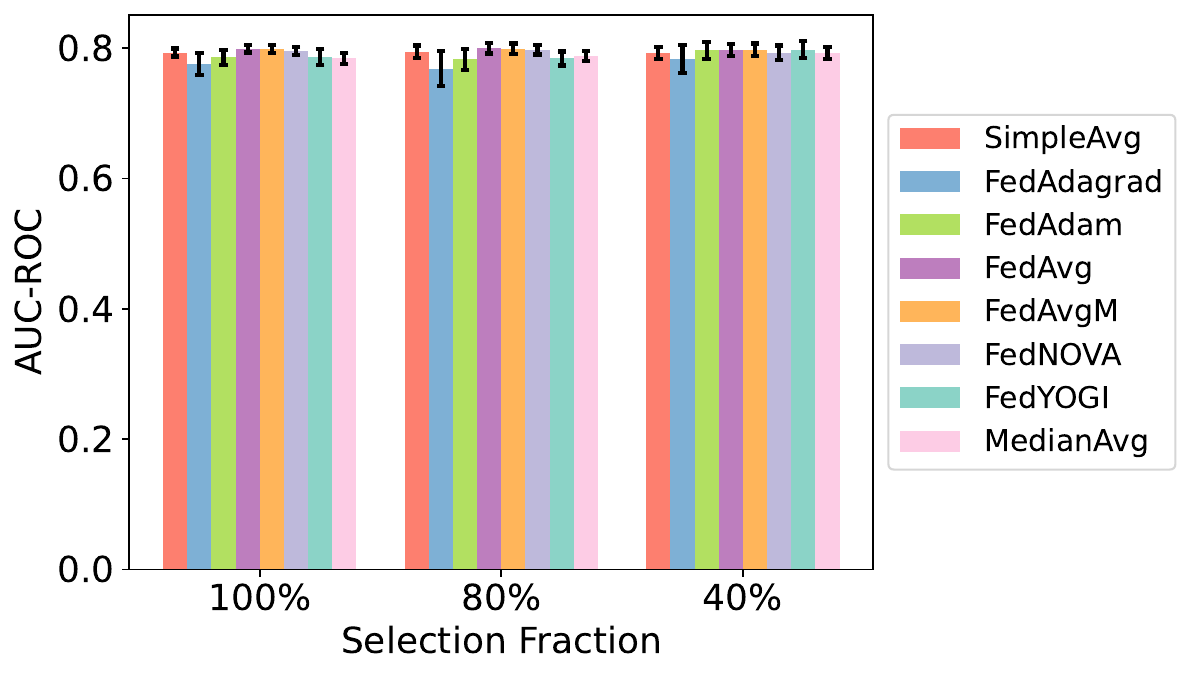}
    \caption{Comparison of Aggregation Algorithms.}
    \label{fig:aggregators}
\end{figure}

To tackle challenges related to statistical heterogeneity and the presence of non-iid data, we evaluate eight popular aggregation algorithms within the FL framework. Fig.~\ref{fig:aggregators} demonstrates the efficacy of the utilized aggregators regarding the AUC-ROC score. Different selection fractions alongside random sampling were employed to test the generalization of our results. From the results, FedAvg and FedAvgM outperform other aggregation algorithms, achieving higher AUC-ROC scores. This suggests that there is no universal 'fit-for-all' solution among aggregation algorithms. In real world applications, we can initially utilize baseline aggregators such as FedAvg. Subsequently, depending on the performance outcomes and specific requirements, exploration and experimentation with alternative aggregation algorithms may be warranted.

\subsection{Client Selection Mechanisms}
\begin{figure}[b!]
    \centering    
    \includegraphics[width=.9\columnwidth]{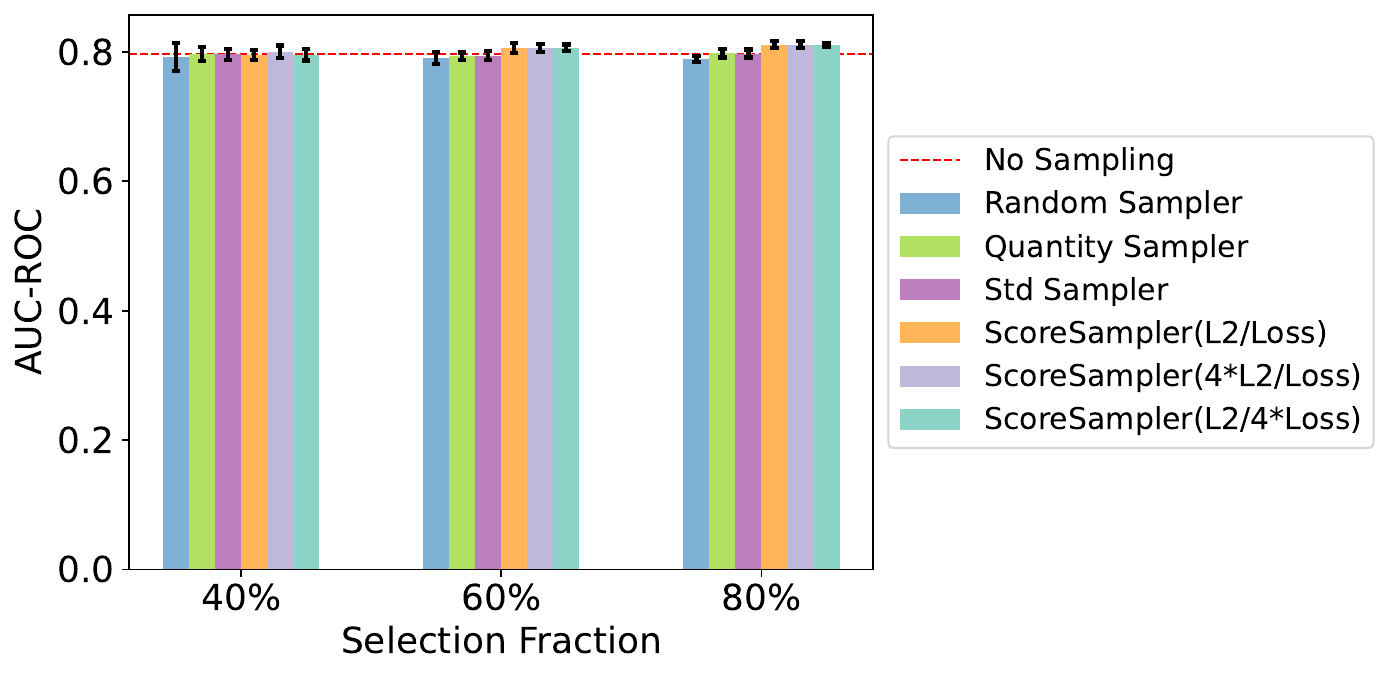}
    \caption{Comparison of Client Selection Meachanisms.}
    \label{fig:selectors}
\end{figure}
\begin{figure}[t!]
    \centering
    \begin{subfigure}[b]{0.8\textwidth}
        \centering
        \includegraphics[width=\textwidth]{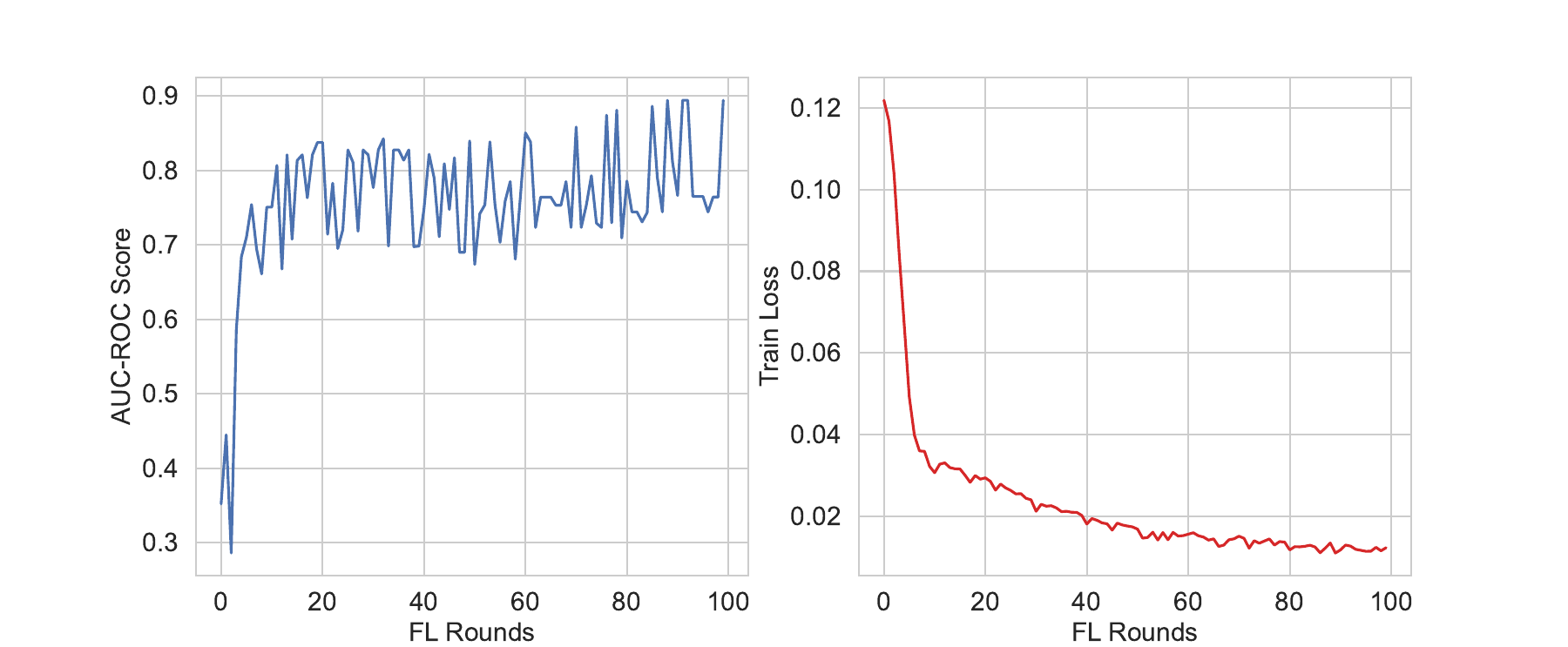}
        \caption{RandomSampler}
        \label{fig:subfig_a}
    \end{subfigure}
    
    \begin{subfigure}[b]{0.8\textwidth}
        \centering
        \includegraphics[width=\textwidth]{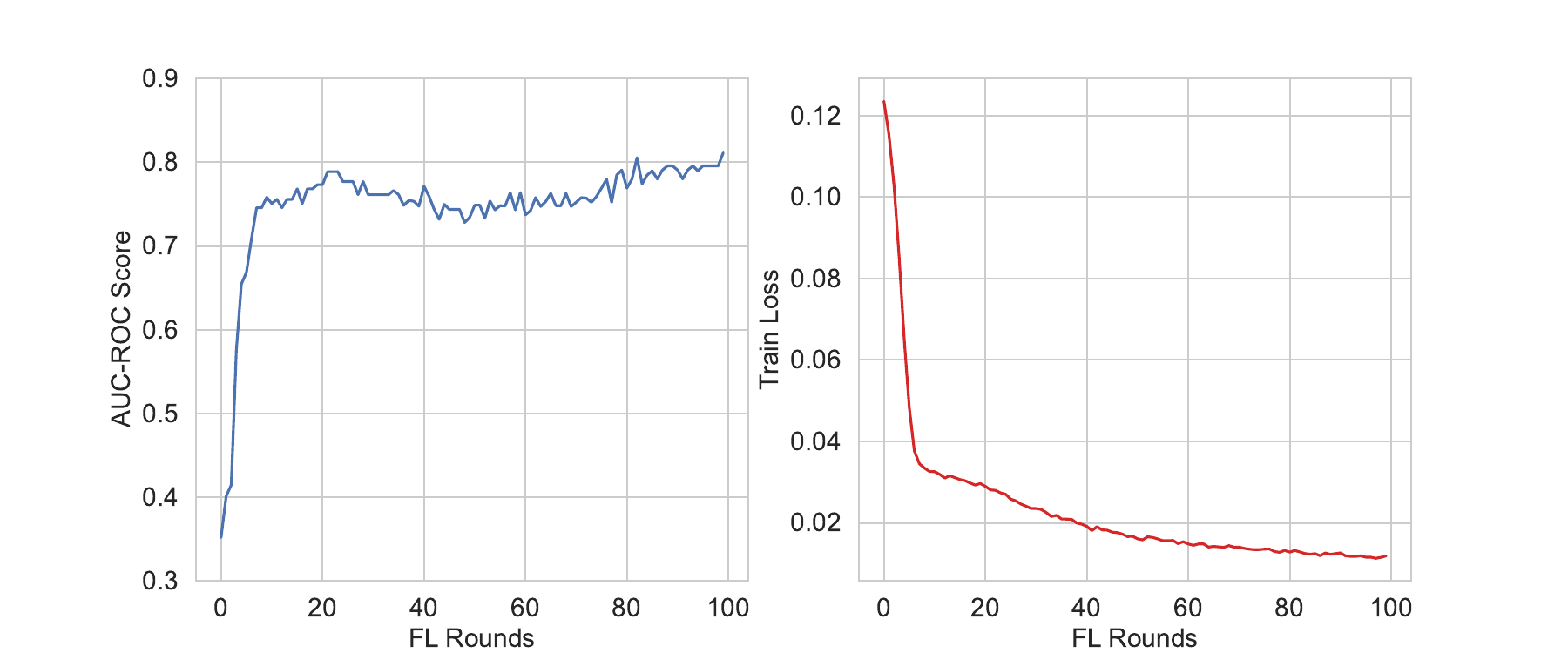}
        \caption{ScoreSampler}
        \label{fig:subfig_b}
    \end{subfigure}
    \caption{Comparing convergence of different client selection mechanisms.}
    \label{fig:samplers_convergence}
\end{figure}

Another significant aspect of this work is designing an effective client selection mechanism that can improve predictive accuracy and offer learning stability and robust results. In Fig.~\ref{fig:selectors}, we present the results from the experimental evaluation of the utilized client selection algorithms for different selection fractions. The ScoreSampler, which is proposed in this work based on the contribution of each client, not only outperforms the rest client selection techniques but also the experiment where no client selection is utilized. The most promising result though is depicted in Fig.~\ref{fig:samplers_convergence}. It is evident that the SocreSampler has a more consistent and robust convergence curve compared to the baseline random sampling. This leads to the conclusion that our proposed contribution score based sampler not only offers higher predictive accuracy, but also ensures a more stable training. That can be attributed to the fact that the ScoreSampler can effectively exclude highly divergent models from aggregation, which, in other techniques like random sampling, could lead to performance degradation.

\subsection{Fully Homomorphic Encryption}
\begin{figure}[ht!]
    \centering    
    \includegraphics[width=.9\columnwidth]{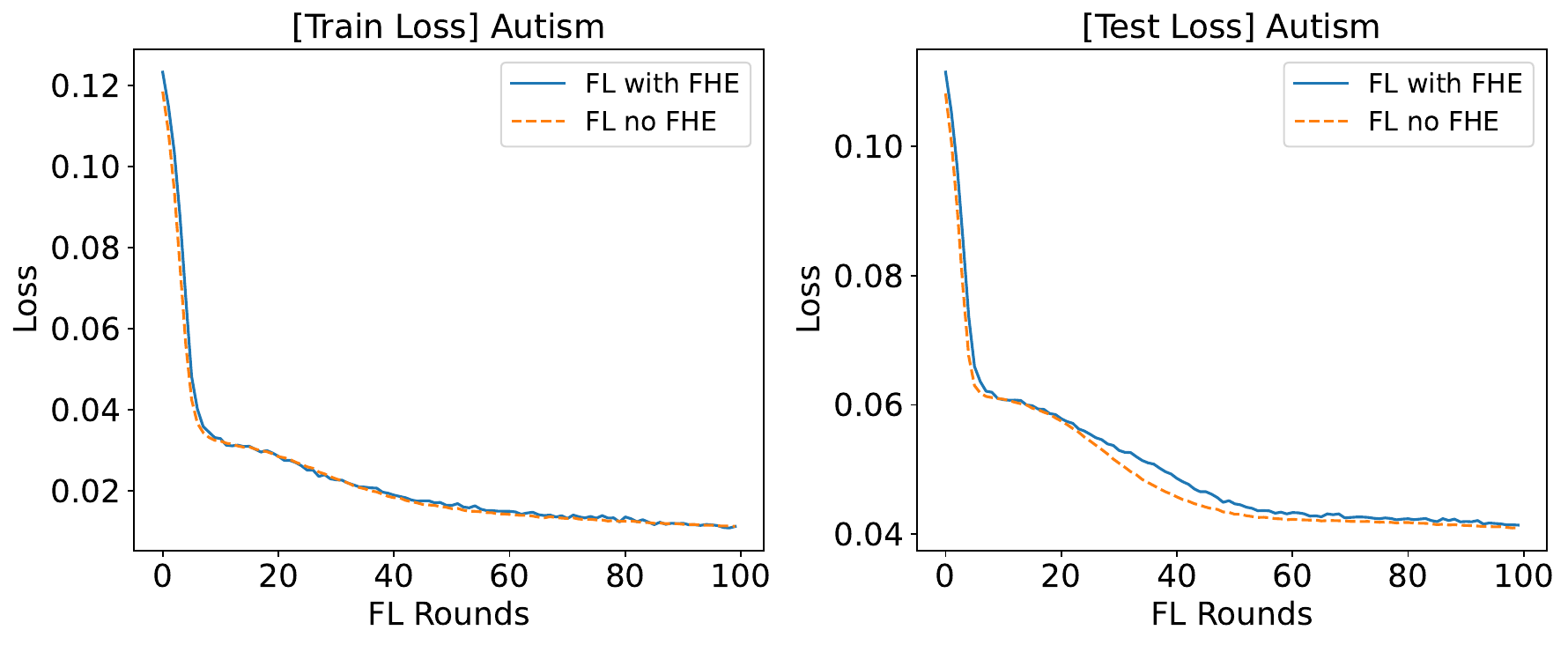}
    \caption{Evaluating Impact of FHE on AutoEncoder's convergence.}
    \label{fig:flfhe}
\end{figure}

As a last experiment, we assess the impact of FHE on the federated process. As shown in Fig. \ref{fig:flfhe} FHE has no significant impact nor on the convergence curve neither on the predictive performance of the ML model using the FedAvg aggregation. This result leads to the conclusion that FHE can be used as a privacy-enhancing technique for the FL during the learnable weights aggregation step, while not significantly affect the overall performance and efficiency.

\section{Conclusion \& Future Work}
\label{conclusion}
Developing an effective framework for identification of ASD in children, while preserving the privacy of sensitive data can be very challenging. In this work, we proposed a Federated Learning solution that integrates semi-supervised Anomaly detection to identify outliers that are prone to develop communication deficiencies. We proved that our proposed solution can have similar results to other works from the literature that incorporate centralized solutions, while ensuring user privacy. Also, the anomaly detection approach showcased very promising results, especially for datasets that the ground truth labels are not available for all the cases. Furthermore, we introduced a novel client selection algorithm that can improve overall efficacy of the framework. We also proved that simple aggregators lead to better predictions, thus there is no need for specialized aggregators that are proposed in other works. Finally, we employed FHE that enabled securing exchanged gradients between clients and server, while at the same time maintaining the performance.

Currently, our proposed framework is operational in a production setting with five clinicians. In this real-world application, the model undergoes monthly updates, allowing for continuous improvement and adaptation to emerging data patterns. This practical implementation underscores the viability and effectiveness of our Federated Learning solution in addressing the complex challenge of ASD identification in children.

Future directions of this work could follow two discrete paths. The first one focuses on the published datasets. More ML techniques, e.g. classification or clustering, could be employed and evaluate their performance under different learning settings. The other path of future research steps focuses on the developed framework itself. For instance, different preprocessing techniques can be utilized in order to evaluate their impact on the overall performance of the framework regarding both predictive accuracy and the data privacy.

\begin{acks}
This work has been partially funded by the \grantsponsor{GS501100001809}{European Union's Horizon 2020 research and innovation programme}{https://terminet-h2020.eu/open-call-winners/} under the TERMINET project with Grant No.: \grantnum{}{957406}.
\end{acks}

\bibliographystyle{ACM-Reference-Format}
\bibliography{main}

\end{document}